\documentstyle[sprocl,epsf]{article}

\def\be{\begin{equation}}
\def\ee{\end{equation}}
\def\bea{\begin{eqnarray}}
\def\eea{\end{eqnarray}}

\begin{document}

\title{
$B\rightarrow X_s\tau^+\tau^-$ in a CP spontaneously broken\\
two Higgs doublet model}

\author{
Chao-Shang Huang \footnote{Presenting author} and
Shou Hua Zhu 
}

\address{
Institute of Theoretical Physics, Academia Sinica,
      P.O.Box 2735,
Beijing 100080, P.R.China}


\maketitle\abstracts{ 
The differential branching ratio, CP asymmetry and lepton polarization $P_N$ for
 $B\rightarrow X_s\tau^+\tau^-$ in a CP spotaneously broken two Higgs doublet model are computed.
It is shown that contributions of neutral Higgs bosons  to the decay are
quite significant when $\tan\beta$ is large and $P_N$ can reach five persent.}

The origin of the CP violation has been one of main issues
in high energy physics.  
The measurements of electric dipole moments of the neutron and 
electron and the matter-antimatter asymmetry in the universe
indicate that one needs new sources of CP violation in addition
to the CP violation come from CKM matrix, which has been one 
of motivations to search new theoretical models beyond the 
standard model (SM).

The minimal extension of the SM is to enlarge the Higgs sectors
of the SM. It has been shown that if one adheres to the
natural flavor conservation (NFC) in the Higgs sector, then
a minimum of three Higgs doublets are necessary in order to have
spontaneous CP violations \cite{new2}. However, the constraint
can be evaded if one allows the real and image parts of $\phi_1^+ 
\phi_2$ have different self-couplings and adds a linear term of Re($\phi^+_1\phi_2$)
in the Higgs potential (see below Eq. (\ref{eq2}) ).
Then, one can construct a CP spontaneously broken (and $Z_2$-symmetry softly broken) two Higgs 
doublet (2HDM), which is the minimal 
among the extensions of the SM that provide  new source of CP violation.

Rare decays $B\rightarrow X_sl^+l^-(l=e,\mu)$ have been 
extensively investigated in both SM and the beyond \cite{gsw,ex1}. 
The inclusive decay $B\rightarrow X_s\tau^+
\tau^-$ has also been investigated in the SM, the model II 2HDM and
SUSY models  with and without
including the contributions of NHB \cite{Dai,hz}.
In this note we investigate the inclusive decay $B\rightarrow X_s\tau^+\tau^-$
 with emphasis on CP violation effect in a CP spontaniously broken 
2HDM in which the up-type quarks get masses from
Yukawa couplings to the one Higgs doublet $H_2$ 
and down-type quarks and leptons get masses from Yukawa
couplings to the another Higgs doublet $H_1$.
 The contributions from exchanging neutral Higgs
bosons now is enhanced roughly by a factor of $tg^2\beta$ and can compete with
those from exchanging $\gamma,~Z$ when $tg\beta$ is large enough.
 We shall be interested in the large $\tan\beta$ limit in this note.

Consider two complex $Y=1$ $SU(2)_w$ doublet scalar fields, $\phi_1$ and
$\phi_2$. The Higgs potential which spontaneously breaks $SU(2)\times U(1)$
down to $U(1)_{EM}$ can be written in the following form \cite{hz,vend}:
\begin{eqnarray}
V(\phi_1,\phi_2) &=&
\sum_{i=1,2} [m_i^2 \phi_i^+ \phi_i +\lambda_i  (\phi_i^+ \phi_i)^2] 
+ m_3^2 Re(\phi_1^+\phi_2) \nonumber\\
&&+ \lambda_3 [ (\phi_1^+ \phi_1)(\phi_2^+ \phi_2) ]
+ \lambda_4
 [ \mbox{Re}(\phi_1^+ \phi_2)]^2
+ \lambda_5
 [ \mbox{Im}(\phi_1^+ \phi_2) ]^2
 \label{eq2}
\end{eqnarray}
Hermiticity requires that all parameters are real so that
the potential is CP conservative.
It is easy to see that the minimum of the potential is at
$<\phi_1>=
v_1
$
and
$<\phi_2>=
v_2 e^{i\xi},
$
%
thus breaking $SU(2)\times U(1)$
down to $U(1)_{EM}$ and
simutaneously breaking CP, as desired. 

In the case of large $\tan\beta$,
if we neglect all terms proptional to $c_\beta$ and taken $s_\beta=1$ in the 
mass-squared matrix of neutral Higgs,
one would get  that one of the Higgs boson mass is 0, obviously which is conflict with
current experiments. So instead, below we will discuss a special case in which
a analytical solution of the mass matrix can be obtained.

Assuming $4\lambda_1c^2_{\beta}=\lambda_4-\lambda_5$ and neglecting
other terms proptional to $c_\beta$ in the mass-squared matrix,
one obtains the masses of neutral Higgs bosons
\begin{eqnarray}
m_{H^0_1}^2=2(\lambda_4- \lambda_5) v^2 s^2_{\psi},~~~
m_{H^0_2}^2=2(\lambda_4- \lambda_5) v^2 c^2_{\psi},~~~ 
m_{H^0_3}^2=4 (\lambda_2+\lambda_3) v^2 
\label{eq101}
\end{eqnarray}
with the mixing angle $\psi = \frac{\xi}{2}$.

Then it is straightforward to obtain the couplings of neutral Higgs to fermions 
\begin{eqnarray}
H^0_1 \bar f f:\   -\frac{i g m_f}{2 m_w c_{\beta}} 
(-s_{\xi/2}+ i c_{\xi/2} \gamma_5);
\ \ H^0_2 \bar f f: \ 
   -\frac{i g m_f}{ 2 m_w c_{\beta}} (c_{\xi/2}+ i s_{\xi/2} \gamma_5), 
 \label{eq111}
\end{eqnarray}
where $f$ represents down-type quarks and leptons. The coupling of $H_3^0$ to f is
not enhenced by tan$\beta$ and will not be given explicitly. The
couplings of the charged Higgs bosons to fermions are the same as
those in the model II 2HDM.
This is in contrary with the model III~\cite{new4} in which the couplings
of the charged Higgs to fermions are quite different from model II.
It is easy to see from Eq. (\ref{eq111}) that the contributions
come from exchanging NHB is proportional to 
$\sqrt{2} G_F s_{\xi/2} c_{\xi/2}  m_f^2/\cos^2\beta$, so that 
we have the constraint
$\sqrt{|\sin \xi|}\tan\beta < 50
$
from the neutron EDM, and the constraint from the electron EDM
is not stronger this. 
The constraints on $\tan\beta$ due to
effects arising from the charged Higgs are the same as those in the
model II and can be found in ref.~\cite{15,Dai}.

The transition rate for
$b\rightarrow s\tau^+\tau^-$ can be computed in the framework of the QCD 
corrected effective weak Hamiltonian
\begin{equation}\label{ham}
H_{eff}=\frac{4G_F}{\sqrt{2}}V_{tb}V^*_{ts}(\sum_{i=1}^{10}C_i(\mu)O_i(\mu)
+\sum_{i=1}^{10}C_{Q_i}(\mu)Q_i(\mu))
\end{equation}
where $O_i(i=1,\cdots ,10)$ is the same as that given in the ref.\cite{gsw}, $Q_i$'s
come from exchanging the neutral Higgs bosons and are defined in Ref. 
\cite{Dai}. 

The coefficients $C_i$'s at $\mu=m_W$ have been given in the ref.\cite{gsw} and $C_{Q_i}$'s are
(neglecting the $O(tg\beta)$ term)
\begin{eqnarray}
C_{Q_1}(m_W)&=&\frac{m_bm_{\tau}tg^2\beta x_t}{2 sin^2\theta_W}
\{
\sum_{i=H_1,H_2} \frac{ A_{i}}{m_{i}^2} (f_1 B_{i}+f_2 E_i) \},
\nonumber \\
C_{Q_2}(m_W)&=&\frac{m_bm_{\tau}tg^2\beta x_t}{2 sin^2\theta_W}
\{
\sum_{i=H_1,H_2} \frac{ D_{i}}{m_{i}^2} (f_1 B_{i}+f_2 E_i) \},
\nonumber \\
C_{Q_3}(m_W)&=&\frac{m_be^2}{m_{\tau}g_s^2}(C_{Q_1}(m_W)+C_{Q_2}(m_W)),
\nonumber 
\\
C_{Q_4}(m_W)&=&\frac{m_be^2}{m_{\tau}g_s^2}(C_{Q_1}(m_W)-C_{Q_2}(m_W)), \nonumber
\\
C_{Q_i}(m_W)&=&0, ~~~~i=5,\cdots, 10
\label{eq1}
\end{eqnarray}
where
\begin{eqnarray}
A_{H_1}&=&-s_{\xi/2},~~~  D_{H_1}=i c_{\xi/2},~~~
A_{H_2}= c_{\xi/2},~~~ D_{H_2}=i s_{\xi/2},
\nonumber \\
B_{H_1}&=&\frac{ i c_{\xi/2}-s_{\xi/2}}{2},  ~~~~~~~~
B_{H_2}=\frac{c_{\xi/2}+i s_{\xi/2}}{2} 
\nonumber \\
E_{H_1}&=& \frac{1}{2} (-s_{\xi/2} c_1+ c_{\xi/2} c_2),~~~~ 
E_{H_2}=\frac{1}{2} (c_{\xi/2} c_1+ s_{\xi/2} c_2) 
\nonumber \\
c_1&=& -x_{H^\pm}+\frac{c_\xi}{2 s^2_{\xi/2}} x_{H_1} (c_\xi+i s_\xi)
+\frac{1}{2 s^2_{\xi/2}} x_{H_1}
\nonumber \\
c_2&=& i[- x_{H^\pm}+\frac{c_{\xi/2} x_{H_1}}{ s_{\xi/2}} (s_\xi-i c_\xi)].
\end{eqnarray}

The QCD corrections to coefficients $C_i$ and $C_{Q_i}$ can be incooperated
in the standard way by using the renormalization group equations~\cite{Dai}.
The explict expressions of  the 
invariant dilepton mass distribution, CP asymmetries $A_{CP}^i$ (i=1,2)
in branching ratio
and in forward-backward (F-B) asymmetry and normal polarization of lepton $P_N$ 
for $B\rightarrow X_s \tau^+\tau^-$ can be found in ref. \cite{hz}. 

The following parameters have been used in the numerical calculations:
$
m_t=175Gev,~m_b=5.0Gev,~m_c=1.6Gev,~m_{\tau}=1.77Gev,~\eta=1.724,
$
$
m_{H_1}=100 Gev, ~m_{H^\pm}=200 Gev.
$

From the numerical results, the following conclusions can be drawn. 
A. The contributions of NHB to the differential branching ratio
$d\Gamma/ds$ are significant when $\tan\beta$ is not smaller than 30 and the
masses of NHB are in the reasonable region. For tan$\beta$=30, $d\Gamma/ds$ is enhenced
by a factor of 6 compared to SM.
B. The direct CP violation in branching ratio $A_{CP}^1$ is of order $10^{-4}$ which is hard 
to be measured. 
C. The direct CP violation in F-B $A_{CP}^2$  can  reach 
about a few percents and is strongly dependent of the CP violation
phase $\xi$ and comes mainly from exchanging NHBs, as expected.
D. The CP-violating polarization $P_N$ is also strongly dependent of
the CP violation phase $\xi$ and can be as large as 5\% for some
values of $\xi$, which should be within the luminosity reach of
coming B factord comes mainly from NHB contributions in the most of range
of $\xi$. So it is possible to discriminate the
model from the other 2HDMs by measuring the CP-violated observables
such as $A_{CP}^2$, $P_N$ if the nature chooses large $\tan\beta$.
\footnote{
This research was supported in part by the National Nature Science
Foundation of China and the post doctoral foundation
of China. S.H. Zhu gratefully acknowledges
the support of  K.C. Wong Education Foundation, Hong Kong.
}




\end{document}